\documentclass[twocolumn,showpacs,showkeys,preprintnumbers]{revtex4-1}
\usepackage{amssymb}

\usepackage{amsfonts}

\usepackage{latexsym}

\usepackage{dcolumn}

\usepackage{amsmath}

\usepackage{graphicx}

\usepackage{psfrag,pstricks}

\begin{document}

\title{Back-action evading measurement of the collective mode of a Bose-Einstein condensate}

\author{M. Fani$^{1}$}

\author{A. Dalafi$^{1}$ }
\email{a\_dalafi@sbu.ac.ir}
\affiliation{$^{1}$ Laser and Plasma Research Institute, Shahid Beheshti University, Tehran 19839-69411, Iran}

\date{\today}

\begin{abstract}
In this paper, we investigate theoretically the back-action evading measurement of the collective mode of an interacting atomic Bose-Einstein condensate (BEC) trapped in an optical cavity which is driven coherently by a pump laser with a modulated amplitude. It is shown that for a specified kind of amplitude modulation of the driving laser, one can measure a generalized quadrature of the collective mode of the BEC indirectly through the output cavity field with a negligible back-action noise in the good-cavity limit. Nevertheless, the on-resonance added noise of measurement is suppressed below the standard quantum limit (SQL) even in the bad cavity limit. Moreover, the measurement precision can be controlled through the \textit{s}-wave scattering frequency of atomic collisions.
\end{abstract}

\maketitle

\section{Introduction}
Since the advent of quantum mechanics the measurement of systems without destroying the quantum effects has been a very important and challenging issue \cite{Braginsky1992}. As is well-known, the presence of decoherency and different kinds of noises limit the precision of measurements \cite{Clerk2010}. Generally, in every continuous measurement of a quantum system there are three noise sources, the thermal, the imprecision and the back-action noise \cite{Clerk2010}. The classical thermal noise can be suppressed by cooling the system to cryogenic temperatures, but the balance between the imprecision and the back-action noises imposes a standard quantum limit (SQL) on the  measurement precision which originally arises from the Heisenberg uncertainty relation \cite{Clerk2010}. 

However, in many cases such as gravitational wave detection \cite{Braginsky1980,Corbitt2004}, quantum information protocols \cite{Brooks,Kozlov,Appel2009}, optical atomic clocks \cite{Ludlow2015} etc., a measurement precision beyond the SQL is essential. Therefore, the development of high-precision quantum measurement is crucial for the future of quantum technologies. For this reason, in the last decades many efforts have been made to beat the SQL and various methods have been proposed which can be divided into three general categories \cite{Miao}. The First one is based on the generation of correlations between the imprecision and the back-action noise by modification of the input field for example by using the squeezed input light \cite{Kimble2001,Khalili2009}. The Second category contains methods which modify the dynamics of systems to cancel the back-action noise such as coherent quantum noise cancellation (CQNC) through the introduction of a negative mass oscillator \cite{Tsang2010,Khalili2018,Buchmann2016,Moller2017} or suppress the noises and amplify the signal simultaneously through parametric modulation of the system parameters \cite{Levitan2016,Motazedi2019}. The last one which is discussed in this paper is the quantum non-demolition (QND) measurement \cite{Braginsky1980,Eckert2008,Sewell2013}. It is based on the measurement of a special observable which is not affected by the back-action noise at the price of injection of a large back-action noise to its conjugate operator. 

\begin{figure}[htbp]
	\centering
	\includegraphics[width=7.5cm]{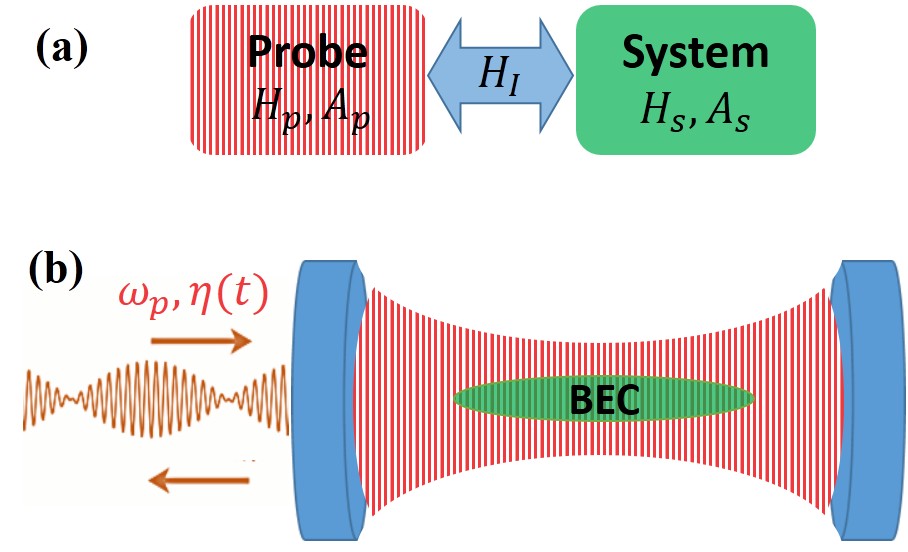}
	\caption{(Color online) (a) Schematic diagram for an indirect measurement, (b) Schematic of a one-dimensional Bose Einstein condensate trapped inside the optical lattice of an optical cavity  which is pumped coherently by a laser with a modulated amplitude. Here, the Bogoliubov mode of the BEC (the quantum system indicated by the green color) is measured indirectly by the optical mode of the cavity which plays the role of the quantum probe system (indicated by the vertical red stripes).}
	\label{Fig1}
\end{figure}

The QND measurement which is also called the back-action evasion measurement was first introduced by Braginsky \textit{et al.} \cite{Braginsky1980} to the purpose of the gravitational wave detection and then it was applied to other cases such as spin measurement \cite{Takahashi1999}, atomic  magnetometery \cite{Shah2010}, single photon detection \cite{Nogues1999}, and measurements based on optomechanical cavities \cite{Heidmann1997,Jacobs1994}. In the QND measurement, in order to evade the back-action noise, one measures a so-called QND variable, $ A_s $, of a quantum system with Hamiltonian $H_s$ which should satisfy the condition, $ [A_s(t),A_s(t')] = 0 $ in the ideal situation. The most famous continuous QND variables are those which are conserved during the free evolution, i.e., they satisfy the equation  $ i \hbar \frac{\partial A_s}{\partial t} + [A_s , H_s] = 0 $ \cite{Braginsky1980,Tsang2012}. In addition, the QND measurement is an indirect measurement so that the system variable is measured indirectly by an observable $A_p$  of another quantum system, the so-called probe system, with Hamiltonian $H_p$ which has been coupled to the system through an interaction Hamiltonian, $ H_I $ as has been demonstrated in Fig.\ref{Fig1}(a). \\

In order to have an ideal QND measurement, $ H_I$ has to fulfill the following requirements \cite{Scully}: (i) $\frac{\partial H_I}{\partial A_s}\neq 0$, (ii) $[A_s , H_I] = 0$, and (iii) $ [A_p , H_I]\neq 0 $. Based on the first condition, the interaction Hamiltonian should be directly dependent on the system variable. The second condition ensures that the coupling of the system to the probe dose not affect the dynamics of the system variable (back-action evasion criterion). The third condition requires that the interaction Hamiltonian affects the dynamics of $A_p$ so that the information corresponding to $A_s$ is transformed to  $A_p$ which is measured directly.


In order to realize the above-mentioned high precision quantum measurement methods, optomechanical systems have been always considered as appropriate candidates. For example, there are several schemes for force and mass sensors which have been proposed and implemented in different optomechanical systems based on the CQNC method \cite{Tsang2010,Buchmann2016,Moller2017,Khalili2018} where the radiation pressure noise is canceled by destructive interference through an anti-noise path. Besides, a back-action evading measurement of a cantilever has been also proposed in a standard optomechanical cavity by a modulated input field which also leads to the conditional squeezing of the mechanical oscillator \cite{Clerk2008}. This scheme which is called the two-tone back-action evading measurement has been realized experimentally \cite{Shomroni2019,Hertzberg2010}. Furthermore, the problem of parametric instability of the present scheme has been investigated theoretically and experimentally \cite{Shomroni2019b,Steinke2013}. After this proposal, modulated optomechanical systems have attracted a lot of attention \cite{Mari2009,Malz2016,Aranas2016,Motazedi2018,Motazedi2019a}. It can also be generalized to the back-action evading measurement of the collective quadrature of two mechanical oscillators \cite{Woolley2013}. In addition, it has been found that the parametric modulation of the spring constant can lead to the simultaneous noise suppression and signal amplification \cite{Levitan2016,Motazedi2019}.

On the other hand, it has been shown \cite{Kanamoto2010} that a Bose-Einstein condensate (BEC) of atoms trapped in an optical cavity is equivalent to an optomechanical system where the Bogoliubov mode of the BEC (the collective excitation of condensate) plays the role of an effective mechanical oscillator . Because of the robustness, low decoherency, high controlability and the possibility to hybridize with other systems such as mechanical resonators, the BEC is an important system in observing the macroscopic quantum effects \cite{Dalafi2018,Dalafi2017a,Byrnes2012,Andrews1997,Nagy2009}. Besides, in ultracold quantum gases the atomic collisions lead to a controllable non-linearity \cite{Dalafi2014,Dalafi2013a,Dalafi2017,Dalafi2015}. Furthermore, a BEC can be considered as a tool to measure the acceleration (in particular the gravity acceleration) and thus acts as a force sensor via measuring the Bloch oscillations in an optical lattice \cite{Debs2011,Georges2017,Rosi2014}.

Based on the the above-mentioned investigations, we have been motivated to generalize the two-tone back-action evading method proposed in Ref.\cite{Clerk2008} to an effective optomechanical system consisting of a BEC in order to do a QND measurement of the collective excitation of the BEC. For this purpose, we consider an atomic BEC trapped in an effectively one-dimensional trap inside an optical cavity which is pumped by an external laser with a modulated amplitude. In the dispersive regime of atom-field interaction and under the Bogoliubov approximation, the collective mode of the BEC behaves as an effective mechanical oscillator which couples to the radiation pressure of the cavity. Here, the collective mode of the BEC is considered as the quantum system to be measured while the radiation pressure of the cavity field is used as the probe system (Fig.\ref{Fig1}). In this way, we introduce a QND observable for the BEC and determine the circumstances which the QND conditions are met. Then, it is shown that the the collective excitation of the BEC can be measured at the output of the cavity with a precision under the SQL without disturbing the BEC dynamics. 

The remainder of this paper is organized as follows, in section \ref{secmodel} we introduce the physical model of the system. In section \ref{dynamics} we study the dynamics of the system by the input-output method. Section \ref{bogol} is dedicated to the calculation of the Bogoliubov mode spectra. The output field spectrum is also calculated in section \ref{output}. In addition, we discuss all the results in section \ref{results} and finally the summary and conclusion are given in section \ref{conclusion}.

\section{Physical Model}\label{secmodel}

As shown schematically in figure \ref{Fig1}(b), we consider a cigar-shaped BEC consisting of $N$ identical two-level atoms with mass $m_a$ and transition frequency $\omega_{a}$ trapped inside an optical cavity with length $ L $. The cavity is pumped coherently by a laser light with the time-dependent amplitude $\eta(t) $ and the central frequency $ \omega_p $. In the dispersive regime where $ \omega_p $ is far detuned from the atomic  resonance, i.e., when $ \Delta_a = \omega_p - \omega_a $ exceeds the atomic linewidth by orders of magnitude, the excited electronic state of the atoms can be adiabatically eliminated and spontaneous emission can be neglected \cite{Maschler2004}. Besides, it is assumed that the atoms are trapped in a cylindrical dipole trap where its longitudinal trapping frequency is much smaller than its transverse frequency. In this way, the dynamics of atoms can be described within an effective one-dimensional model by quantizing the atomic motional degree of freedom along the $x$ axis \cite{Ritter2009}. Therefore, the Hamiltonian of the system is given by \cite{Dalafi2016,Nagy2009}
\begin{equation}\label{hamilton}
H = \hbar \omega_c a^{\dagger} a + i \hbar \eta (t) \left( a e^{i \omega_p t} - a^{\dagger} e^{- i \omega_p t} \right) + H_{BEC} + H_{diss},
\end{equation} 
where $ H_{BEC}$ is the BEC Hamiltonian which is given by \cite{Maschler2004},
\begin{eqnarray}\label{bechamilton}
H_{BEC} = \int\limits_{-L/2}^{L/2} dx \Psi^{\dagger} (x)  \left[ - \frac{\hbar^2}{2 m_a} \frac{d^2}{dx^2} + \hbar U_0 a^{\dagger} a \cos^2{(k x)} \right.  \nonumber \\
+ \left.  \frac{1}{2} U_s \Psi^{\dagger}(x) \Psi(x) \right]  \Psi(x)  ,   
\end{eqnarray}
and $ H_{diss}$ denotes the dissipative processes corresponding to both the cavity and the BEC \cite{Dalafi2015}. Here, we have assumed that a single-mode optical field has been formed inside the cavity with the bare frequency $ \omega_c $  whose annihilation (creation) operator is $ a (a^{\dagger}) $. In the equation (\ref{bechamilton}), $\Psi(x)$ is the quantum field operator of the atoms in the framework of the second quantization formalism, $k=\omega_p/c$ is the wave number of the optical field and $ U_0 = g_0^2 / \Delta_a $ is the optical lattice depth per photon which results from the dispersive atom-cavity interaction with the vacuum Rabi frequency $g_{0}$. Besides, we have also considered an atom-atom scattering term with strength $ U_s =  4 \pi \hbar^2 a_s / m_a $ where $ a_s $ is the s-wave scattering length.

When the optical lattice is not very deep, i.e., if the condition $ U_0 \langle a^{\dagger} a \rangle \le 10 \omega_R  $ is satisfied (here $ \omega_R = \hbar^2 k^2 /2m_a $ is the recoil frequency of the condensate atoms) and in the Bogoliubov approximation \cite{Nagy2009}, the matter wave operator can be expanded as a single mode field \cite{Nagy2009,Dalafi2016,Dalafi2013}
\begin{equation}\label{Psi}
\Psi(x)=\sqrt{\frac{N}{L}} + \sqrt{\frac{2}{L}} \cos{(2 k x)} c,
\end{equation}
where the first term denotes the condensate phase which is written as a c-number in the Bogoliubov approximation and $ c $ is the annihilation operator of the first excited mode.
Substituting equation (\ref{Psi}) in the Hamiltonian of (\ref{bechamilton}), the Hamiltonian (\ref{hamilton}) can be written in the rotating frame at frequency $\omega_p $  as follows
\begin{subequations}
	\begin{eqnarray}\label{H}
	&&	H =- \hbar \Delta_c a^{\dagger} a + i \hbar \eta (t) ( a  - a^{\dagger} ) + H_{c} + H_{diss} ,\\ \label{Hc}
	&&	H_c = \hbar \Omega_c c^{\dagger} c + \hbar g a^{\dagger} a (c + c^{\dagger}) + \frac{1}{4} \hbar \omega_{sw} (c^2 + {c^{\dagger}}^2) ,
	\end{eqnarray}
\end{subequations}
where $\Delta_c = \omega_p - \omega_c - \frac{1}{2} N U_0 $ is the effective detuning, $ \Omega_c = 4 \omega_R + \omega_{sw} $ is the effective frequency of mode $c $. The second term of the Hamiltonian (\ref{Hc}) shows the coupling of the cavity mode to the excited mode of the BEC with the effective coupling strength $g =  U_0 \sqrt{2 N} /4$ which is analogous to an optomechanical coupling. The last term of (\ref{Hc}) arises from the atom-atom interaction with the \textit{s}-wave scattering frequency $\omega_{sw}=  \frac{8 \pi \hbar a_s N}{m_a L w^2}$, where $w$ is the optical mode waist. 

To diagonalize the BEC Hamiltonian (\ref{Hc}), we use the following Bogoliubov transformation
\begin{subequations}
	\begin{eqnarray}
	&& b = \frac{1}{2 \chi} [(\chi^2 + 1) c + (\chi^2 - 1) c^{\dagger} ],\\
	&& b^{\dagger} = \frac{1}{2 \chi} [(\chi^2 - 1) c + (\chi^2 + 1) c^{\dagger} ],
	\end{eqnarray}
\end{subequations}
where $\chi = (\Omega_+ / \Omega_-)^{1/4} $ and $ \Omega_\pm = \Omega_c \pm \frac{1}{2} \omega_{sw} $. In this way, the total Hamiltonian of the system in terms of the Bogoliubov mode, $b$ is given by 
\begin{eqnarray}\label{Hb}
H &=&- \hbar \Delta_c a^{\dagger} a + \hbar \omega_m b^{\dagger} b + \hbar G a^{\dagger} a (b + b^{\dagger}) \nonumber \\
&+& i \hbar \eta (t) ( a  - a^{\dagger}  ) + H_{diss},
\end{eqnarray}
where $ \omega_m = \sqrt{\Omega_- \Omega_+} $ is the effective Bogoliubov frequency, and $ G = g /\chi $ is the coupling strength. It should be noticed that the Hamiltonian of equation (\ref{Hb}) is quit similar to a standard optomechanical system, but with the time-dependent pump amplitude.

\section{Dynamics of the system}\label{dynamics}

To describe the dynamics of the system, we use the Hamiltonian of equation (\ref{Hb}) to obtain the following  Heisenberg-Langevin equations
\begin{subequations}
	\begin{eqnarray}
	&&\dot{a} =i \Delta_c a - \frac{\kappa}{2} a + \eta(t) - i G a (b + b^{\dagger}) + \sqrt{\kappa} a_{in},\label{LHea}\\
	&&\dot{b} = - i \omega_m b - \frac{\gamma}{2} b - iG a^{\dagger} a + \sqrt{\gamma} b_{in}, \label{LHeb}
	\end{eqnarray}
\end{subequations}
where $ \kappa$ and $ \gamma $ are, respectively, the decay rates of the cavity and the Bogoliubov modes. In addition, $ a_{in} $ and $ b_{in} $ are the input noise operators with thermal noise correlation functions,
\begin{subequations} \label{correlation}
	\begin{eqnarray}
	&&	\left\langle a^{\dagger}_{in} (t) a_{in}(t') \right\rangle = \bar{n}^{th}_a \delta (t-t')\\
	&& \left\langle a_{in} (t) a^{\dagger}_{in}(t') \right\rangle = (\bar{n}^{th}_a +1) \delta (t-t'), \\
	&&	\left\langle b^{\dagger}_{in} (t) b_{in}(t') \right\rangle = \bar{n}^{th}_b \delta (t-t'),\\
	&&	\left\langle b_{in} (t) b^{\dagger}_{in}(t') \right\rangle = (\bar{n}^{th}_b +1) \delta (t-t'),
	\end{eqnarray}
\end{subequations}
where $ \bar{n}^{th}_a = (e^{\hbar \omega_c / k_B T}-1)^{-1} $ and $ \bar{n}^{th}_b = (e^{\hbar \omega_m / k_B T}-1)^{-1} $ are the thermal excitations of cavity and Bogoliubov mode, respectively. Since $ \hbar \omega_c \gg k_B T $ the thermal excitation of the cavity is negligible.

Since equations (\ref{LHea}) and (\ref{LHeb}) are nonlinear, they cannot be solved analytically. Thus, to proceed, we write the optical and atomic fields as the sum of their classical mean values and quantum fluctuations as $a = \alpha (t) + \delta a$ and $b = \beta (t) + \delta b$, respectively. It should be noticed that due to the time dependence of the pump amplitude, the mean fields, $\alpha $ and $ \beta $ are also time dependent and they satisfy the following equations of motion
\begin{subequations}\label{mean}
	\begin{eqnarray}\label{mean1}
	&&\dot{\alpha} = (i\Delta'_c - \kappa/2)\alpha+ \eta (t),\\ \label{mean2}
	&&\dot{\beta} = (-i \omega_m - \gamma/2) \beta - i G |\alpha|^2,
	\end{eqnarray}
\end{subequations}
where $ \Delta'_c = \Delta_c - G (\beta + \beta^*)$ is the effective detuning. In addition, the linearized equations of motion corresponding to the quantum fluctuations are given by
\begin{subequations}\label{eqmdelta}
	\begin{eqnarray}
	&&\delta\dot{a} = (i \Delta'_c - \kappa/2) \delta a - i G \alpha (\delta b + \delta b^{\dagger}) + \sqrt{\kappa} a_{in},\label{eqmdelta a}\\
	&&\delta\dot{b} = (-i\omega_m - \gamma/2) \delta b - i G (\alpha^* \delta a + \alpha \delta a^{\dagger}) + \sqrt{\gamma} b_{in}.\label{eqmdelta b}
	\end{eqnarray}
\end{subequations}
Notice that in equations (\ref{eqmdelta a}) and (\ref{eqmdelta b}) we have neglected the terms which contain the product of two or more fluctuating operators. 

Now, by defining the optical quadratures $X = \frac{1}{\sqrt{2}} (\delta a + \delta a^{\dagger})$ and $Y = \frac{-i}{\sqrt{2}} (\delta a - \delta a^{\dagger})$, and the generalized Bogoliubov quadratures which rotate at arbitrary frequency $\Omega$,
\begin{subequations}
	\begin{eqnarray}
	Q_{\Omega} = \frac{1}{\sqrt{2}} (\delta b e^{i \Omega t} + \delta b^{\dagger} e^{-i \Omega t}), \label{QOmega}\\
	P_{\Omega} = \frac{-i}{\sqrt{2}} (\delta b e^{i \Omega t} - \delta b^{\dagger} e^{-i \Omega t}),\label{POmega}
	\end{eqnarray}
\end{subequations}
the following set of linearized equations will be obtained by using the equations (\ref{eqmdelta a}) and (\ref{eqmdelta b}) and taking the time derivative of the generalized Bogoliubov quadratures,
\begin{subequations}\label{eqmX}
	\begin{eqnarray}\label{eqmXa}
	\dot{X} &=&- \Delta'_c Y - \frac{\kappa}{2} X + \sqrt{\kappa} X_{in} \nonumber\\ 
	&+& 2 G {\mathop{\rm Im}\nolimits} \alpha (t) (Q_{\Omega} \cos \Omega t + P_{\Omega} \sin \Omega t) 	,\\ \label{eqmXb}
	\dot{Y}&=& \Delta'_c X -  \frac{\kappa}{2} Y + \sqrt{\kappa} Y_{in} \nonumber \\
	&-& 2 G {\mathop{\rm Re}\nolimits} \alpha (Q_{\Omega} \cos \Omega t + P_{\Omega} \sin \Omega t), \\\label{eqmXc}
	\dot{Q}_{\Omega} &=& (\omega_m - \Omega)P_{\Omega} -  \frac{\gamma}{2} Q_{\Omega} + \sqrt{\gamma} Q_{in} \nonumber \\
	&+& 2 G \sin \Omega t ( {\mathop{\rm Re}\nolimits} \alpha X + {\mathop{\rm Im}\nolimits} \alpha Y)	,\\ \label{eqmXd}
	\dot{P}_{\Omega}&=& - (\omega_m - \Omega)Q_{\Omega} - \frac{\gamma}{2} P_{\Omega} + \sqrt{\gamma} P_{in} \nonumber \\
	& -& 2 G \cos \Omega t ( {\mathop{\rm Re}\nolimits} \alpha X + {\mathop{\rm Im}\nolimits} \alpha Y) .
	\end{eqnarray}
\end{subequations}

The above equations are quite general for any time-dependence of the pump amplitude and any $ \Omega$, but since we are interested in the QND measurement on the BEC, we have to choose the parameters $ \Omega $ and $ \eta (t) $, so that the QND conditions are satisfied. Before that, let us remind that the effective Hamiltonian from which the linearized equations (\ref{eqmXa})-(\ref{eqmXd}) are obtained can be written as follows
\begin{subequations}
	\begin{eqnarray}
	&& H_{eff}= H_s +H_p + H_ I +H_{diss} ;\label{Heff}\\ 
	&&H_s =\frac{1}{2} \hbar \omega_m (Q_{\Omega}^2 + P_{\Omega}^2),\label{Hs} \\ 
	&&H_p = -\frac{1}{2} \hbar \Delta'_c (X^2 + Y^2),\\ \label{Hp}
	&&H_I = 2 \hbar G ( Q_{\Omega} \cos\Omega t +  P_{\Omega} \sin \Omega t) ( X {\mathop{\rm Re}\nolimits} \alpha + Y {\mathop{\rm Im}\nolimits} \alpha ). \label{HI}
	\end{eqnarray}
\end{subequations}

Here, the Hamiltonian of the system to be measured, i.e., $H_s$, is that of the Bogoliubov mode of the BEC, $H_p$ is the Hamiltonian of optical mode of the cavity which plays the role of the probe system and $H_I$ is the interaction Hamiltonian between them. In addition, it is obvious that the Heisenberg equations for $Q_{\Omega}(t)$ and $P_{\Omega}(t)$ which lead to the equations (\ref{eqmXc}) and (\ref{eqmXd}) should be considered as $\dot A=\frac{1}{i\hbar}[A,H_{eff}]+\frac{\partial}{\partial t}A$ for $A=Q_{\Omega}, P_{\Omega}$ because of their explicit time dependence.

Now, based on the definitions of (\ref{QOmega}) and (\ref{POmega}) together with the system Hamiltonian of (\ref{Hs}), the only way to have the continuous QND variables, i.e., to fulfill the condition, $ i \hbar \frac{\partial A_s}{\partial t} + [A_s , H_s] = 0 $, is that $ \Omega = \omega_m $. In this way, the QND variables of the BEC can be defined as $ Q \equiv Q_{\omega_m} $ and $ P \equiv P_{\omega_m} $. In fact, these quadratures rotate at the effective Bogoliubov frequency $ \omega_m $, so measuring one of them does not lead to any back action due to the free evolution. In the rest of this paper, we choose $ Q $ as $ A_s $, i.e., the QND operator of the system which is going to to be measured. By this choice, the first condition of $ H_I $, i.e., the condition (i), is obviously fulfilled by the interaction Hamitonian of (\ref{HI}) because it depends on $Q$. Furthermore, without loss of generality we can assume $ {\mathop{\rm Im}\nolimits} \alpha = 0 $, which is not very restricting and can be achieved by adjusting the pump phase. Therefore, if we choose $ Y $ as the probe operator $ A_P$, then the third QND condition, i.e., the condition (iii), is also satisfied by the interaction Hamiltonian (\ref{HI}).

Nevertheless, by the above-mentioned choices the second QND condition, i.e., the condition (ii), is not still satisfied because $ [Q , H_I] = 2 i\hbar G X \alpha(t) \sin \omega_m t\neq 0 $. However, if the Fourier transform of $ \alpha (t) \sin \omega_m t $ only contains terms with frequencies greater than or equal to $2\omega_m$, and also if the spectral density of the Bogoliubov mode is so narrow that it dose not respond to the frequencies much larger than $ \omega_m $, i.e., if $\gamma\ll\omega_{m}$, then the effect of the above mentioned commutator in the system dynamics will be negligible. Although in this case the measurement is not an ideal QND one, it can be considered as a back-action evading measurement which can surpass the SQL as will be shown in the next sections. One of the simplest forms of $ \alpha(t) $ which satisfies this condition is
\begin{equation}\label{alpha}
\alpha(t) = \alpha_{max} \cos \omega_m t,
\end{equation}
where we will consider it as the optical mean field and will also specify the explicit form of the time modulation of the pump amplitude which leads to this mean-field amplitude in the cavity (see equation. (\ref{etat})). By substituting equation (\ref{alpha}) into the set of equations (\ref{eqmXa})-(\ref{eqmXd}) the Heisenberg-Langevin equations take the following form
\begin{subequations}\label{eqmfinal}
	\begin{eqnarray}
	\dot{X}&=&- \Delta'_c Y - \frac{\kappa}{2} X + \sqrt{\kappa} X_{in},\label{eqmfinala} \\
	\dot{Y}&=& \Delta'_c X -  \frac{\kappa}{2} Y - G \alpha_{max} (1 +\cos (2 \omega_m t) ) Q \nonumber \\
	&-& G \alpha_{max} \sin (2\omega_m t)  P  + \sqrt{\kappa} Y_{in},\\ \label{eqmfinalb}
	\dot{Q}&=&  - \frac{\gamma}{2} Q +  G \alpha_{max} \sin (2 \omega_m t) X  + \sqrt{\gamma} Q_{in},\label{eqmfinalc}\\
	\dot{P}&=& - \frac{\gamma}{2} P -  G  \alpha_{max} (1+\cos (2 \omega_m t)) X +  \sqrt{\gamma} P_{in}.\label{eqmfinald}
	\end{eqnarray}
\end{subequations}
As is seen from the set of equations of (\ref{eqmfinala})-(\ref{eqmfinald}), for $ \Delta'_c=0 $ the back-action noise of measurement which is injected to the quadrature $Q$ through the second term in the right-hand side of equation (\ref{eqmfinalc}) is minimized because in this case, the extra noises corresponding to $Y$ are no longer transferred to $Q$ due to decoupling of $X$ from $Y$.

Using the mean-field equation (\ref{mean1}) and considering the above-mentioned conditions, the pump amplitude $ \eta (t) $ corresponding to the intra cavity field amplitude (\ref{alpha}) can be specified as
\begin{equation}\label{etat}
\eta(t)= \eta_{max} \cos (\omega_m t + \phi),
\end{equation}   
where the amplitude and the phase are, respectively, given by $ \eta_{max} = \alpha _{max}\sqrt {\frac{{{\kappa ^2}}}{4} + \omega _m^2} $ and $\phi  = \arctan(2\omega_{m}/\kappa)$. The expression (\ref{etat}) shows that the pump amplitude should be modulated at the frequency of Bogoliubov mode which means the cavity should be driven by two lasers tuned at the both first sidebands of the cavity, i.e., $ \omega_p \pm \omega_m $, with the same amplitude, $ \eta_{max}/2 $ and with the phase difference, $ \phi $ where $\omega_p$ is determined by the resonance condition $ \Delta'_c=0 $. In addition, putting $\alpha(t)$ of equation (\ref{alpha}) and the Fourier series of $\beta (t)$, i. e., $\beta(t) = \sum_{n} \beta_n e^{i n \omega_m t}$ in equation (\ref{mean2}), we find that the only nonzero Fourier components of $\beta(t)$ are,
\begin{eqnarray}
&&\beta_0 = -\frac{i G \alpha_{max}^2}{2(i \omega_m + \gamma/2)} \approx - G \alpha_{max}^2 / 2 \omega_m ,\\
&&\beta_2 = -\frac{i G \alpha_{max}^2}{4(3 i \omega_m + \gamma/2)} \approx - G \alpha_{max}^2 / 12 \omega_m ,\\
&&\beta_{-2} = -\frac{i G \alpha_{max}^2}{4(-i \omega_m + \gamma/2)} \approx  G \alpha_{max}^2 / 4 \omega_m ,
\end{eqnarray}
where the approximated expressions are valid for $ \gamma \ll \omega_m $, which is compatible with the experimental data. In order to calculate the effects of the back action and the added noise due to the coupling of the BEC to the cavity field, in the next section, we calculate the spectra of the Bogoliubov mode. 

\section{Bogoliubov mode spectra}\label{bogol}
To calculate the spectra, we need to write the equations (\ref{eqmfinala})-(\ref{eqmfinald}) in the frequency domain by using the Fourier transform,
\begin{equation} \label{Fourier}
A(\omega) = \frac{1}{\sqrt{2 \pi}} \int dt A(t) e^{i \omega t},
\end{equation}
for $ A=X,Y,Q,$ and $P$. Thus, the Heisenberg-Langevin equations in the frequency domain are obtained as
\begin{subequations}\label{eqminomega}
	\begin{eqnarray}\label{eqminomega1}
	&&- i \omega X(\omega) =- \frac{\kappa}{2} X(\omega) - \Delta_c Y(\omega) \nonumber \\
	&& \,\,\,\,\,\,\,\, + G \sum_{n=0,\pm 2} \bar{\beta}_n Y(\omega + n \omega_m)  + \sqrt{\kappa} X_{in}(\omega), \\ \label{eqminomega2}
	&&	- i \omega Y(\omega) = - \frac{\kappa}{2} Y(\omega) + \Delta_c X(\omega) - G \sum_{n=0,\pm 2} \bar{\beta}_n X(\omega + n \omega_m) \nonumber \\ 
	&& \,\,\,\,\,\,\,\, - \frac{1}{2} G \alpha_{max} [2 Q(\omega) + Q(\omega+2\omega_m)+Q(\omega-2\omega_m)] \nonumber \\
	&& \,\,\,\,\,\,\,\, + \frac{i}{2} G  \alpha_{max} [P(\omega + 2\omega_m) - P(\omega - 2 \omega_m)]+ \sqrt{\kappa} Y_{in}(\omega),\\ \label{eqminomega3}
	&& - i \omega Q(\omega) =  - \frac{\gamma}{2} Q(\omega)  + \sqrt{\gamma} Q_{in}(\omega),\nonumber \\
	&& \,\,\,\,\,\,\,\, - \frac{i}{2} G \alpha_{max} [X(\omega + 2\omega_m) - X(\omega - 2\omega_m)] \\ \label{eqminomega4}
	&& -i \omega P(\omega) = - \frac{\gamma}{2} P(\omega)  +  \sqrt{\gamma} P_{in}(\omega), \nonumber \\
	&&\,\,\,\,\,\,\,\,\, - \frac{1}{2} G  \alpha_{max} [2X(\omega) + X(\omega + 2\omega_m) + X(\omega - 2 \omega_m) ],
	\end{eqnarray}
\end{subequations}
where we have defined, $ \bar{\beta}_n = \beta_n + \beta^{*}_{-n} $ for $ n=0,2,-2 $.

Since we are interested in the linear response of the Bogoliubov mode, we calculate $ Q(\omega) $ and $ P(\omega) $ to the first order of $ G \alpha_{max} $. For this purpose, it is enough to obtain $ X(\omega) $ to the zeroth order of $ G \alpha_{max} $ as
\begin{equation} \label{X}
X(\omega) = \chi_c (\omega) \sqrt{\kappa} X_{in} (\omega),
\end{equation}
where $ \chi_c (\omega) = (\frac{\kappa}{2} - i \omega)^{-1} $ is the optical susceptibility. Notice that at this level of approximation, the condition $ \Delta'_c = 0 $ is reduced to $ \Delta_c = 0 $. Now, by Substituting Eq. (\ref{X}) in equations (\ref{eqminomega3}) and (\ref{eqminomega4}), the Bogoliubov mode quadratures are obtained to the first order of $ G \alpha_{max} $ as
\begin{subequations}
	\begin{eqnarray}
	Q(\omega) &=& \sqrt{\gamma} \chi_m (\omega) Q_{in} (\omega) \nonumber \\
	&-&  \frac{i}{2} G \alpha_{max} \sqrt{\kappa} \chi_m (\omega)  [\chi_c (\omega+2 \omega_m) X_{in} (\omega+2 \omega_m)  \nonumber \\
	&-&\chi_c (\omega - 2 \omega_m) X_{in} (\omega - 2 \omega_m) ],  \\
	P(\omega) &=& \sqrt{\gamma} \chi_m (\omega) P_{in} (\omega) \nonumber \\
	&-& \frac{1}{2} G \alpha_{max} \sqrt{\kappa} \chi_m (\omega)  [2\chi_c(\omega) X_{in} (\omega) \nonumber \\
	&+& \chi_c (\omega+2 \omega_m) X_{in} (\omega+2 \omega_m) \nonumber \\
	&+&\chi_c (\omega - 2 \omega_m) X_{in} (\omega - 2 \omega_m) ],
	\end{eqnarray}
\end{subequations}
where $ \chi_m(\omega) = (\frac{\gamma}{2} - i \omega)^{-1} $ is the Bogolibov mode susceptibility. Using these equations and the Fourier transform of correlation functions (\ref{correlation}) the Bogoliubov spectra can be evaluated.

The stationary part of the spectrum which is defined as \cite{Malz2016}
\begin{subequations}
	\begin{eqnarray}
	&& S_A(\omega) = \lim_{T \rightarrow \infty} \int_{-T/2}^{T/2} dt S_A(\omega,t); \\
	&& S_A(\omega,t) = \frac{1}{2} \int_{-\infty}^{\infty} d\tau e^{i \omega \tau} \left\langle A(t) A(t+\tau) +  A(t+\tau) A(t) \right\rangle , \nonumber\\
	\end{eqnarray}
\end{subequations}
leads to the following expression for the spectrum of the observable $ Q $ by using Fourier transform (\ref{Fourier}) 
\begin{equation}\label{SQ}
S_Q (\omega) = \frac{\gamma}{2} |\chi_m (\omega)|^2 \left\lbrace 1 + 2 
\bar{n}^{th}_b + 2 n_{bad} (\omega) \right\rbrace ,
\end{equation}
where $n_{bad}(\omega) $ is defined as
\begin{equation}\label{nbadomega}
n_{bad} (\omega) =  \frac{\kappa}{8\gamma} (G \alpha_{max})^2 [|\chi_c (\omega + 2\omega_m)|^2 + |\chi_c (\omega - 2\omega_m)|^2].
\end{equation}
This expression shows the number of quanta in the frequency domain which is added to the spectrum of $ Q $ due to the interaction with the cavity field. Indeed, $ n_{bad}$ corresponds to the back-action noise injected to the $ Q$ spectrum arising from the coupling of the Bogoliubov mode of the BEC to the probe system.  The subscript $bad$ refers to the bad-cavity limit. Notice that in the limit of $ \kappa \ll\omega_m $, i.e., in the good-cavity limit, $n_{bad}$ at resonance is approximated as $ n_{bad} (0) \approx \frac{\kappa (G \alpha_{max})^2}{16 \gamma \omega_m ^2},$ which goes to zero. It shows that the interaction-induced noise can be neglected in this limit and therefore we can do a back-action evading measurement on $Q$. As we will see in the next section, the important advantage of the hybrid system consisting of a BEC in comparison to the bare optomechanical systems is that the effective frequency $ \omega_m $ of the Bogoliubov mode can be controlled by $ \omega_{sw}$ such that $ n_{bad} $ can be decreased by increasing $ \omega_{sw} $.

On the other hand, the spectrum of the conjugate operator, $ P $ is given by 
\begin{equation}
S_P (\omega) = \frac{\gamma}{2} |\chi_m (\omega)|^2 \left\lbrace 1 + 2 
\bar{n}^{th}_b + 2 n_{bad} (\omega) + 2 n_{BA} (\omega) \right\rbrace ,
\end{equation}
where 
\begin{equation}
n_{BA} (\omega) = \frac{\kappa}{2\gamma} (G \alpha_{max})^2 |\chi_c (\omega)|^2 ,
\end{equation}
is the back-action noise added to the observable $ P $. In the good cavity limit, $n_{BA}$ at resonance is approximated as $ n_{BA} (0) \approx \frac{2}{ \kappa \gamma} (G \alpha_{max})^2 $ which is much greater than $n_{bad}(0)$ as is expected from the uncertainty relation.  

As has been mentioned before, the probe operator is the phase quadrature of the cavity field, i.e., $Y$. It means that to measure the QND variable, i.e., $ Q $, one has to do a Homodyne measurement on the phase quadrature of the output field. Thus in order to show the possibility of back-action evading measurement of $ Q $ and obtain the necessary conditions, in the following we calculate the $ S_{Y_{out}}(\omega) $.   

\section {Output Spectrum}\label{output}

The phase quadrature of the output field is given by $ Y_{out} = Y_{in} - \sqrt{\kappa} Y $ \cite{Clerk2010}.
Thus by using the equation (\ref{eqminomega2}) we get 
\begin{eqnarray}
&&Y_{out} (\omega) = [1 - \kappa \chi_c (\omega)] Y_{in}(\omega) \nonumber \\
&&\,\,\,\,\,\,\,\,\, + G \sqrt{\kappa} \chi_c (\omega) \sum_{n=0,\pm 2} \bar{\beta}_n X (\omega + n \omega_m) \nonumber\\
&& \,\,\,\,\,\,\,\,\, + \frac{1}{2} G \alpha_{max} \sqrt{\kappa} \chi_c (\omega) [2 Q(\omega) + Q(\omega + 2\omega_m) \nonumber \\
&&\,\,\,\,\,\,\,\,\, + Q (\omega - 2 \omega_m)] \nonumber \\
&&\,\,\,\,\,\,\,\,\,- \frac{i}{2} G \alpha_{max} \sqrt{\kappa} \chi_c (\omega) [P(\omega + 2\omega_m) - P (\omega - 2 \omega_m)].
\end{eqnarray}
Therefore, the output spectrum can be written as follows
\begin{equation}\label{syout}
S_{Y_{out}} (\omega) = |{\cal G}(\omega)|^2 A(\omega) \Bigg[\frac{1}{2} +  \bar{n}^{th}_b +  n_{add}(\omega)\Bigg],
\end{equation}
where, we have defined
\begin{eqnarray}
&&{\cal G}(\omega) =  \sqrt{\kappa} G \alpha_{max} \chi_c(\omega),\\ \label{Aomega}
&&A(\omega) = \gamma |\chi_m (\omega)|^2 \nonumber \\
&& \,\,\,\,\,\,\,\,\,\,\,\,\,\,\,\,\,+ \frac{\gamma}{2} [|\chi_m(\omega + 2 \omega_m)|^2 + |\chi_m(\omega - 2 \omega_m)|^2]  .
\end{eqnarray}

The function, $ {\cal G}(\omega) $ is called the gain coefficient in some literature \cite{Clerk2008} because the output spectrum can be written as $ S_{Y_{out}} (\omega) = |{\cal G} (\omega)|^2 \{S_Q(\omega) + S_{1}(\omega)\} $ \footnote{$S_{1}(\omega)$ is a part of the spectrum. Since it is complicated and we do not need its explicit form in the following we will not specify it here. }, which shows that the spectrum of the phase quadrature of the output field relates to the $ Q $ spectrum by the coefficient ${\cal G}(\omega)$. Furthermore, the added noise in the equation (\ref{syout}) is given by 
\begin{eqnarray}\label{naddomega}
n_{add}(\omega) &=& \frac{1}{2 |{\cal G}(\omega)|^2 A (\omega)} + n_{bad}(\omega) \nonumber \\
&+& \frac{\gamma n_{BA}(\omega)}{4 A(\omega)} [|\chi_m(\omega + 2 \omega_m)|^2 + |\chi_m(\omega - 2 \omega_m)|^2] \nonumber\\ 
&+& \frac{\kappa}{2 \alpha^2_{max} A(\omega)} \left\lbrace \bar{\beta}_0^2 |\chi_c(\omega)|^2 \right. \nonumber \\
&+& \left.  \bar{\beta}_2 \bar{\beta}_{-2} [|\chi_c(\omega + 2 \omega_m)|^2 + |\chi_c(\omega - 2 \omega_m)|^2] \right\rbrace ,
\end{eqnarray}
which represents  the increase in the number of the optical field quanta due to the measurement. The SQL is determined by $ n_{add} (\omega) = 1/2 $, so that if $ n_{add} (\omega) <1/2 $ it is said that the SQL has been beaten \cite{Caves1980,Clerk2004} and the measurement is an ultra-precision measurement. To find the conditions for beating the SQL we write the on-resonance added noise in the regime of $\gamma \ll \omega_m , \kappa $ as 

\begin{equation}\label{Nadd0}
n_{add}(0) \approx \frac{1}{16 n_{BA}(0)} +\frac{1}{8} \Big(\frac{\kappa^2}{4\omega_m^2 + \kappa^2/4}\Big) n_{BA}(0).
\end{equation}
Since $n_{BA}$ in this expression depends on $\eta_{max}$, we can find the optimum value for the pump amplitude which minimizes the added noise as
\begin{equation}\label{etaopt}
\eta_{max}^{opt} = \left[  \frac{\gamma (\omega_m^2 + \kappa^2/4)\sqrt{4\omega_m^2 + \kappa^2/4}}{2 \sqrt{2} G^2} \right]^{1/2}.
\end{equation}
Besides, the minimum value of the added noise at resonance which occurs at $\eta_{max}=\eta_{max}^{opt}$ is given by
\begin{equation}\label{n-add-min}
n_{add}^{min}(0) \approx \frac{\sqrt{2}}{4} \frac{ \kappa}{\sqrt{\kappa^2 + 16 \omega_m^2}}.
\end{equation}

As is seen from equation (\ref{n-add-min}), the minimum value of the on-resonance added noise is always smaller than $1/2$ for the optimized value of the pump amplitude, i.e., the SQL is beaten anyway. Nevertheless, $n_{add}^{min}(0)$ can be decreased much below the SQL by increasing the effective frequency of the Bogoliubov mode. Since $\omega_{m}=\sqrt{(4\omega_{R}+\frac{1}{2}\omega_{sw})(4\omega_{R}+\frac{3}{2}\omega_{sw})}$, one can decrease $n_{add}^{min}(0)$ by increasing $\omega_{sw}$ which itself can be controlled by the transverse frequency of the optical trap \cite{Morsch}. The interesting point is that even in the bad cavity limit the added noise can be less than the SQL. However, the larger the ratio of $ \omega_m / \kappa $, the smaller the added noise and therefore the more precise the measurement is.

\section{Results and Discussion}\label{results}
In this section, we discuss the results obtained in the previous section based on the experimentally feasible data given in \cite{Ritter2009}. For this purpose, in all the following plots we consider a BEC of $^{87}$Rb consisting of $ N= 5 \times 10^4 $ atoms prepared in the $5\text{S}_{1/2}$ ground state. The atoms have been trapped in a cavity of length $L=178\mu$m with a bare frequency $ \omega_c =  2.41494 \times 10^{15} \text{Hz} $ corresponding to a wavelength of $ \lambda = 780 \text{nm} $ which couples to the atomic D$_2$ transition corresponding to the atomic transition frequency $ \omega_a = 2.41419 \times 10^{15} \text{Hz} $ and coupling strength $ g_0 = 2 \pi \times 14.1 \text{MHz}$. The recoil frequency of the atoms in the optical lattice of the cavity is  $ \omega_R = \frac{\hbar k_c^2}{2 m_a} = 2\pi \times 3.77 \text{kHz}$. In addition, the cavity decay and Bogoliubov damping rates are $ \kappa = 2\pi \times 13 \text{MHz} $ and  $ \gamma = 0.001 \kappa $, respectively.  
\begin{figure}[htbp]
	\centering
	\includegraphics[width=6.5cm]{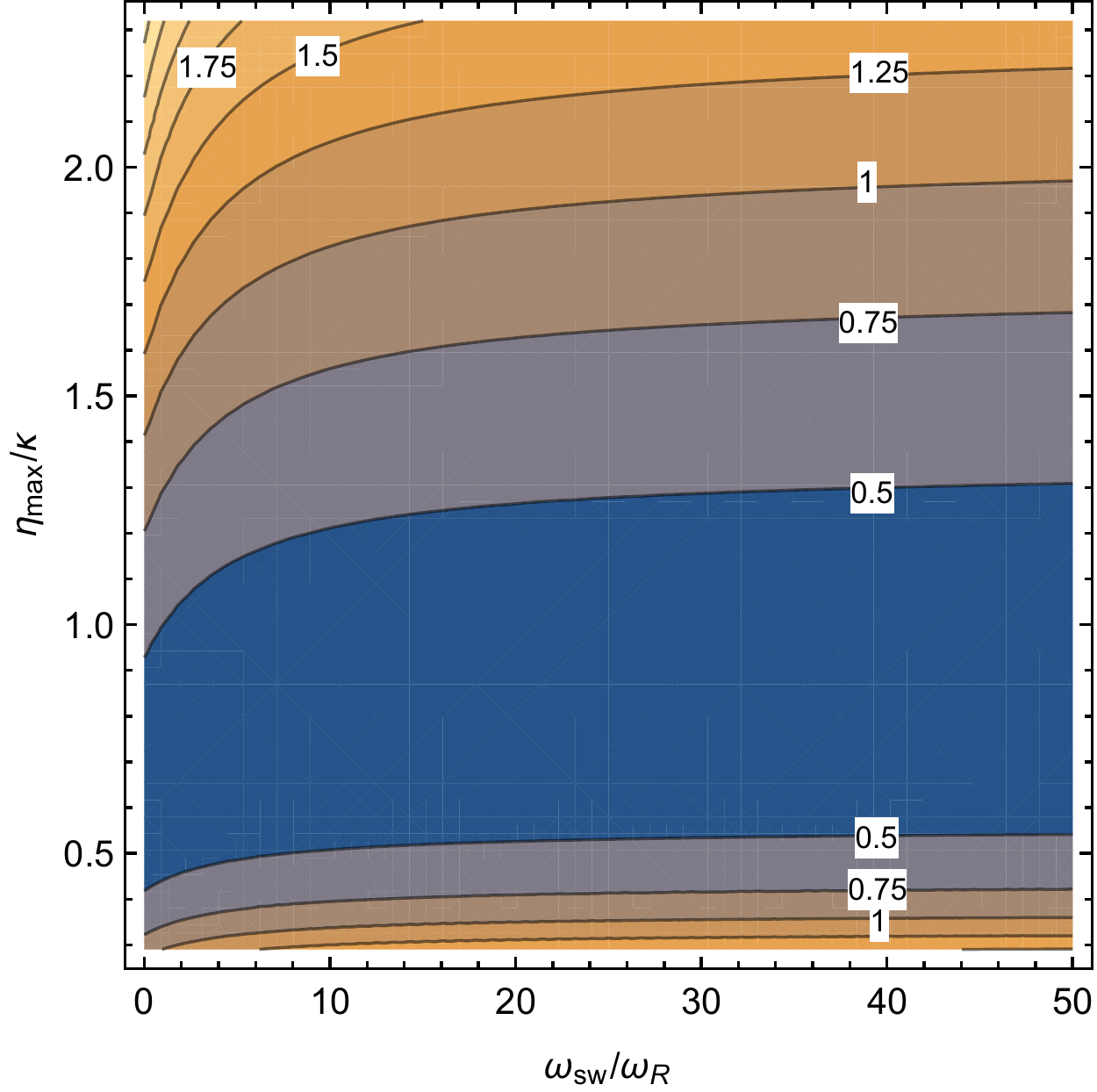}
	\caption{(Color online) Resonant added noise, $n_{add}(0)$ versus \textit{s}-wave scattering frequency $ \omega_{sw}$ and the maximum pump amplitude $\eta_{max}$. The blue region shows the range of the parameters where the standard quantum limit is beaten.}
	\label{fig2}
\end{figure}

In figure (\ref{fig2}) we have demonstrated the contour diagram of the on-resonance added noise $ n_{add}(\omega=0) $ versus the normalized \textit{s}-wave scattering frequency, $\omega_{sw}/\omega_{R} $ and the normalized pump amplitude, $ \eta_{max}/\kappa $. As is seen, there is a region which is shown by the blue color in the diagram that the noise level is below the SQL ($ n_{add}(0) < 1/2 $). This diagram confirms that the $ Q $ quadrature of the BEC can be measured with a precision beyond the SQL by controlling the parameters of the hybrid systems, i.e., $\omega_{sw}$ and $\eta_{max}$.

\begin{figure}[htbp]
	\centering
	\includegraphics[width=7cm]{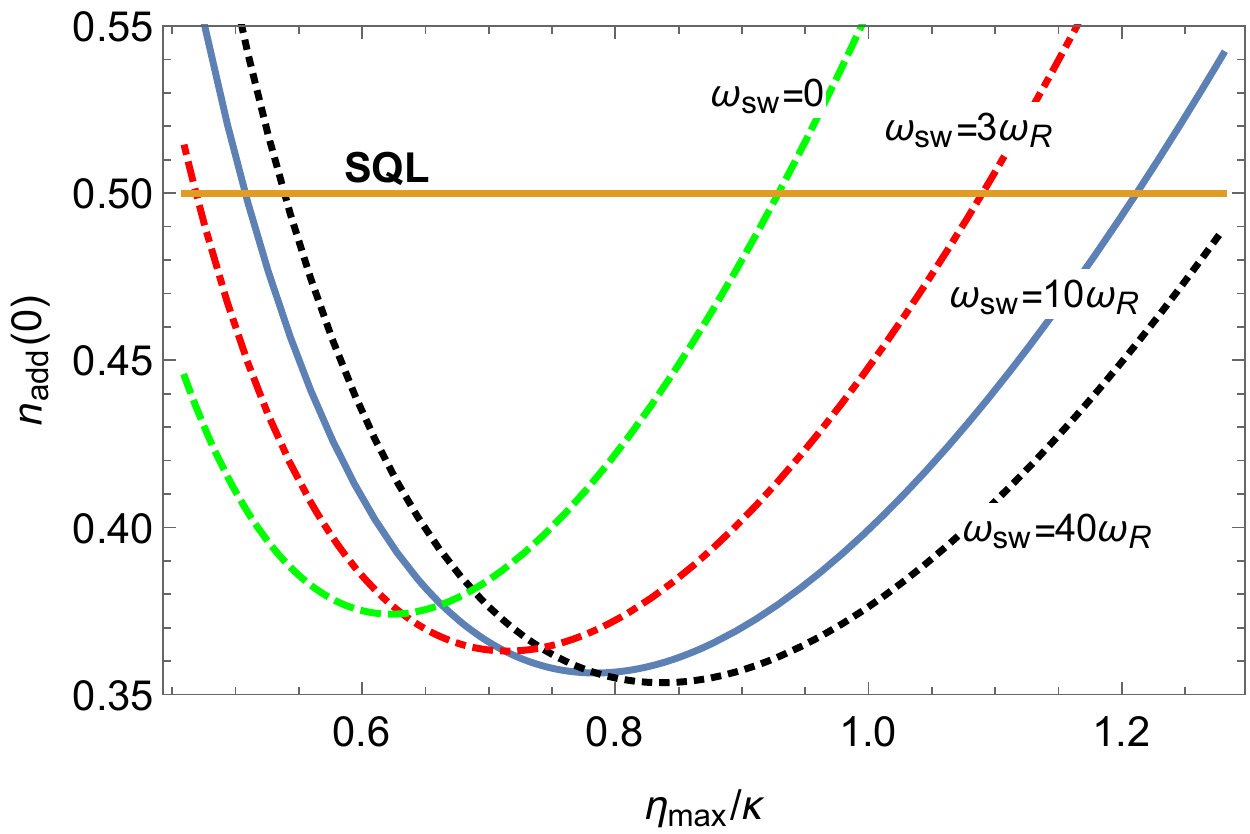}
	\caption{(Color online) The on-resonance added noise versus $ \eta_{max}/\kappa$ for four different values of $\omega_{sw} $.}
	\label{fig3}	
\end{figure}

To see more details, in figure (\ref{fig3}) we have plotted the on-resonance added noise $ n_{add}(0) $ versus $\eta_{max}/\kappa $ in the region that the SQL is beaten based on the data corresponding to the blue region of figure (\ref{fig2}) for four different values of the $ \omega_{sw} $ . As is seen from figure (\ref{fig3}) and as is expected from equation (\ref{etaopt}), for each value of $ \omega_{sw}$ there is an optimum value for the pump amplitude which minimizes the added noise. By increasing the \textit{s}-wave scattering frequency, the effective frequency of the Bogoliubov mode increases which leads to the increase of the optimum value of $\eta_{max}$ based on the equation (\ref{etaopt}) and the decrease of the minimum of the added noise $n_{add}(0)$. In this way, for lager values of $\omega_{sw}$ the minimum of the on-resonance added noise gets lower.

\begin{figure}[htbp]
	\centering
	\includegraphics[width=7cm]{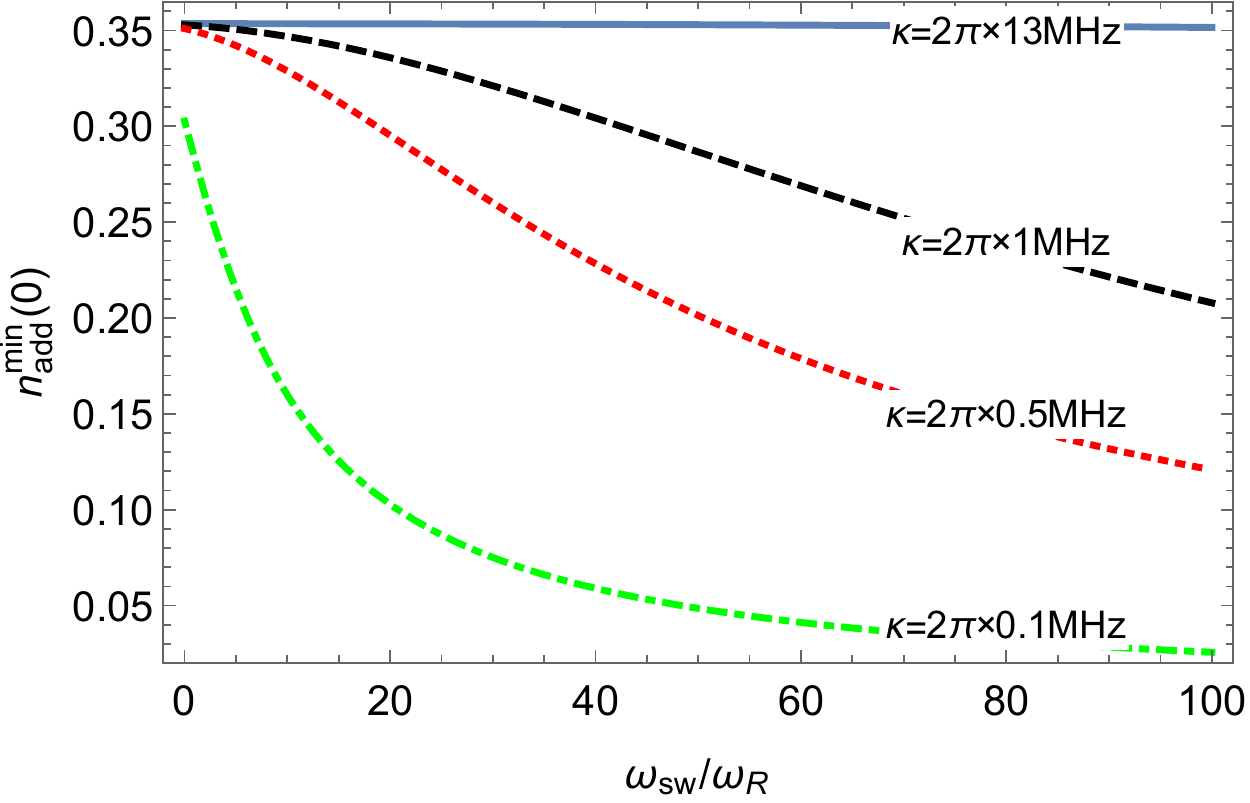}
	\caption{(Color online) The minimum value of the on-resonance added noise versus the \textit{s}-wave scattering frequency for three different values of cavity decay rates.}
	\label{fig4}
\end{figure}

On the other hand, in order to show how much the added noise of measurement in the present hybrid system can be decreased below the SQL, in figure (\ref{fig4}) we have plotted $n_{add}^{min}(0) $, i.e., equation (\ref{n-add-min}), versus the \textit{s}-wave scattering frequency for four different values of the cavity decay rates. It is obviously seen that the minimum of the added noise decreases with the increase of $\omega_{sw}$ for any value of $\kappa$. Nevertheless, the important point is that for the lower cavity decay rates the decrease of $n_{add}^{min}(0) $ is more severe versus $\omega_{sw}$. For example, if the cavity decay rate is of the order of $0.1 \text{MHz}$ the minimum of the added noise can be decreased as low as $0.01$ for sufficiently large values of the \textit{s}-wave scattering frequency. It is because of the fact that the increase of $\omega_{sw}$ makes the effective frequency of the Bogoliubov mode increase and consequently causes the system to go to the good cavity regime.

It should be emphasized that it is one of the most important advantages of the hybrid optomechanical systems consisting of BEC in comparison to the bare ones, that one can control the effective frequency of the atomic mode while the mechanical oscillators in the bare optomechanical systems have fixed natural frequencies which cannot be changed after fabrication. Therefore, the controllability of such hybrid systems gives us the possibility to change the system regime from the bad cavity limit to the good cavity limit through manipulation of the \textit{s}-wave scattering frequency. 

\begin{figure}[htbp]
	\centering
	\includegraphics[width=7cm]{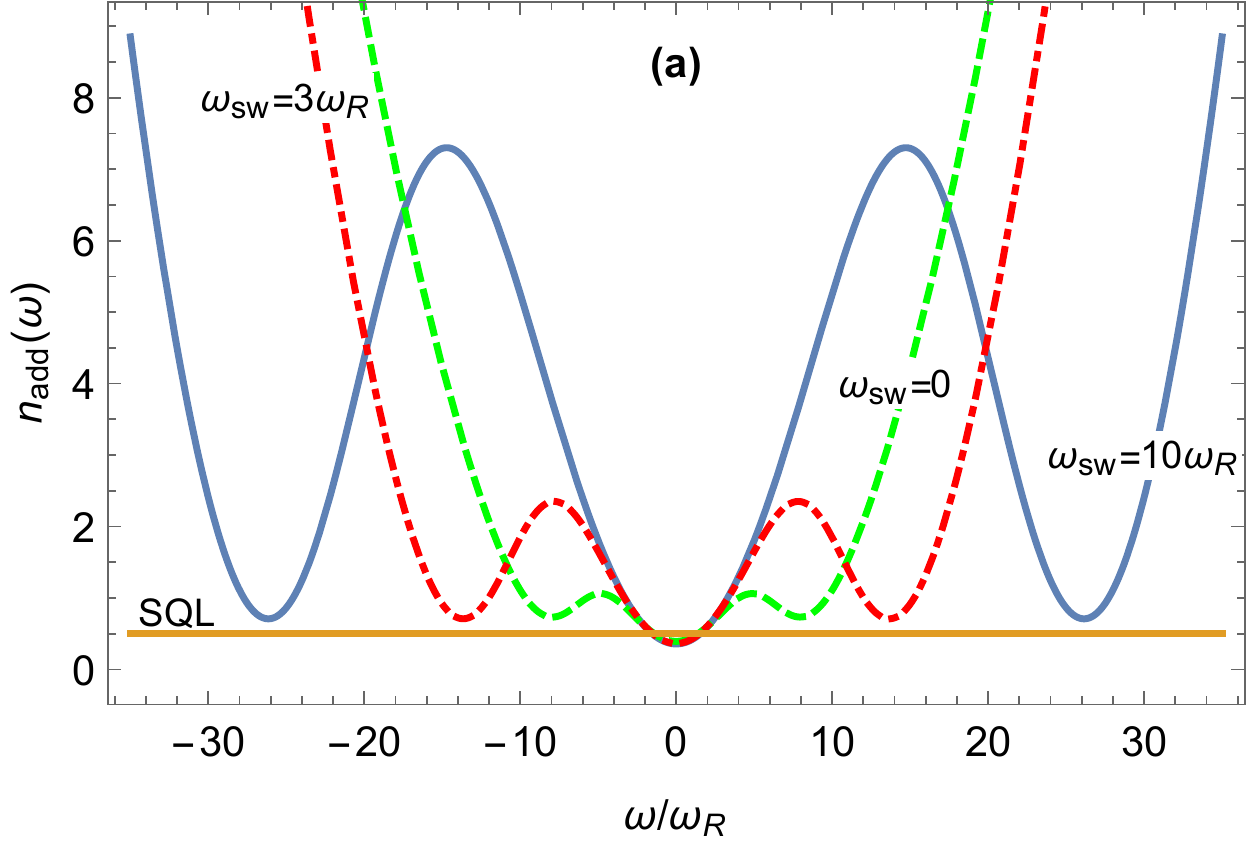}
	\includegraphics[width=7cm]{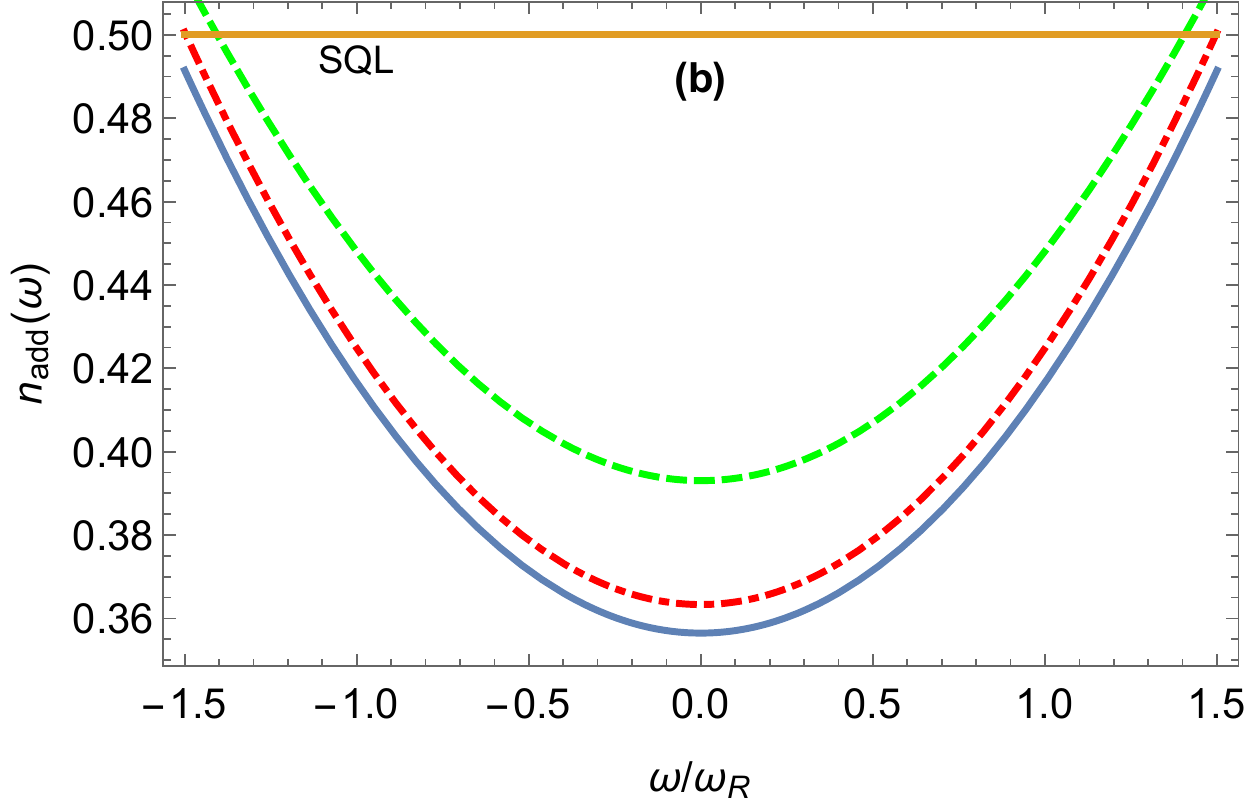}
	\caption{(Color online) (a) The added noise, $n_{add}(\omega)$ for three different values of $ \omega_{sw} $ for the optimum values of the pump amplitude, $ \eta_{max}^{opt} $, (b) The same diagram in a small range near the resonance.  }
	\label{fig5}
\end{figure}

Furthermore, in order to show the frequency dependence of the added noise of measurement, in figure (\ref{fig5}) we have plotted $ n_{add}(\omega)$ for three different values of the \textit{s}-wave scattering frequency under the condition of $\eta_{max}=\eta_{max}^{opt}(\omega_{sw})$ where the on-resonance added noise is minimized. Here, the green dashed curve corresponds to $\omega_{sw}=0$ with $\eta_{max}^{opt}=0.655\kappa$, the red dashed-dotted curve corresponds to  $\omega_{sw}=3\omega_{R}$ with $\eta_{max}^{opt}=0.730\kappa$, and the blue solid curve corresponds to $\omega_{sw}=10\omega_{R}$ with $\eta_{max}^{opt}=0.789\kappa$. As is seen from figure (\ref{fig5}a), the absolute minima of all the curves occur at $\omega=0$ where the SQL is beaten. For clarity, in figure (\ref{fig5}b) the behavior of the added noise has been demonstrated in a narrow interval around $\omega=0$. As is expected, the absolute minimum of the on-resonance added noise is decreased by increasing the \textit{s}-wave scattering frequency. Besides, the frequency window that the noise level is below the SQL becomes a little bit wider.

On the other hand, as is seen from figure (\ref{fig5}a), there are two local minima at $\omega\approx\pm 2\omega_{m}$ for each curve where the added noise gets small, but nevertheless the SQL is not surpassed. It is because of the fact that the dominant term of $n_{add}(\omega)$ is the first term of equation (\ref{naddomega}) which corresponds to the contribution of the imprecision noise. Since the mentioned term is proportional to the inverse of the coefficient $A(\omega)$, it is minimized where $A(\omega)$ is maximized. Based on equation (\ref{Aomega}), the coefficient $A(\omega)$ which is the response function of the Bogoliubov mode has one large central peak at $ \omega = 0 $ and two smaller peaks at $ \pm 2\omega_m $. Therefore, the absolute minimum of the imprecision noise occurs at $\omega=0$ while it has larger minima at $ \pm 2\omega_m $. In addition, the absolut minimum of the imprecision noise at $\omega=0$, becomes equal to the contribution of the back-action noise, i.e., the three other terms in equation (\ref{naddomega}). That is why the SQL is beaten just at the absolute minimum of the added, i.e., at $\omega=0$.

Finally, it should be noticed that there is a one-to-one correspondence between the amount of splitting between the normal modes of the transmitted field of the cavity and the \textit{s}-wave scattering frequency of the atomic collisions as has been shown in Ref.\cite{Dalafi2016}. In fact, by measuring the frequency splitting of the two peaks of the phase noise power spectrum which is experimentally feasible by the homodyne measurement of the light reflected by the cavity, one can estimate the value of the \textit{s}-wave scattering frequency of the atoms. In this way, the effective frequency of the Bogoliubov mode can be measured in terms of the transverse frequency of the optical trap of the BEC which can be calibrated based on the the amount of splitting between the normal modes of the transmitted field of the cavity. So, by knowing the effective frequency of the Bogoliubov mode, one can obtain the optimum value of the pump amplitude as well as the minimum value of the on-resonance added noise necessary for a back action evading measurement through equations (\ref{etaopt}) and (\ref{n-add-min}), respectively.

\section{Conclusion}\label{conclusion}
In this paper, by studying the dynamics of an atomic BEC trapped in an optical cavity which is driven by a coherent field with a time-dependent amplitude, we have investigated the two-tone back-action evading measurement scheme for a collective mode of the BEC, the so-called Bogoliubov mode, which interacts with the radiation pressure of the cavity just like the mechanical oscillator in a bare optomechanical system. In this scheme, the quantum system which is going to be measured is the Bogoliubov mode while the radiation pressure of the cavity plays the role of the probe system. We have shown that if the cavity is driven at the two side bands of the hybrid optomechanical system under the resonance condition with equal amplitudes and with a specified phase difference, then in the output of the cavity field a generalized Q-quadrature of the collective mode of the BEC rotating at the effective frequency of the Bogoliubov mode can be measured with the least back-action noise. Although the mentioned quadrature fulfills the requirements of a QND variable but the measurement is not an ideal QND one except for the extreme regime of the good cavity limit where $\kappa/\omega_{m}$ goes to zero. Nevertheless, it can be considered as a back-action evading measurement which can surpass the SQL even in the bad cavity limit.

One of the most important advantages of the hybrid optomechanical systems consisting of BEC in comparison to the bare ones is that the effective frequency of the Bogoliubov mode is controllable while the mechanical oscillators in the bare optomechanical systems have fixed natural frequencies which cannot be changed after fabrication. Therefore, the controllability of such hybrid systems gives us the possibility to change the system regime from the bad to the good cavity limit through manipulation of the \textit{s}-wave scattering frequency. Since the \textit{s}-wave scattering frequency is controllable through the transverse frequency of the optical trap of the BEC, the added noise of measurement can be reduced effectively through the manipulation of the transverse optical trap.

In the present scheme, the information corresponding to the QND variable of the collective mode of the BEC can be read by the phase quadrature of the output field of the cavity without disturbing the dynamics of the BEC. Specifically, if the modulation amplitude of the input laser is fixed at the optimum value which is determined by the \textit{s}-wave scattering frequency, the back-action of measurement on the collective mode of the BEC can be reduced much below the SQL.

\bigskip


\end{document}